\documentclass[aps,prc,twocolumn,nofootinbib,floatfix,superscriptaddress ]{revtex4-2}

\usepackage{graphicx}
\usepackage{dcolumn}
\usepackage{bm}

\usepackage[english]{babel}

\usepackage{multirow}
\usepackage{amssymb}
\usepackage{amsmath,bbm}
\usepackage{newtxtext,newtxmath}
\usepackage[figuresright]{rotating}
\usepackage{capt-of}
\usepackage{xcolor}
\usepackage{booktabs}
\usepackage{siunitx}
\usepackage{placeins}
\usepackage[hidelinks]{hyperref}
\usepackage{newunicodechar}
\usepackage{orcidlink}

\usepackage[autolanguage]{numprint}

\newcommand{\diff}[3][]{\frac{\partial^{#1}{#2}}{\partial{#3}^{#1}}}
\newcommand{\intdif}[3]{\int_{#1}^{#2}\!\!\!\textnormal{d}{#3}}
\newcommand{\expo}[1]{\textnormal{e}^{#1}}

\newcommand{\ii}{\textnormal{i}}
\newcolumntype{d}[1]{D{.}{.}{#1}}

\newcommand{\nuc}[2]{{}^{#1}\textnormal{#2}}
\renewcommand{\H}[1]{\nuc{1}{H}}

\begin{document}
\title{
Fine-tunings in radiative {\boldmath$\alpha$}-particle capture on {\boldmath$^{12}$}C at astrophysical energies}

\author{Ulf-G.~Mei{\ss}ner\orcidlink{0000-0003-1254-442X}}
\email{meissner@hiskp.uni-bonn.de}
\affiliation{Helmholtz-Institut f\"{u}r Strahlen- und Kernphysik and Bethe Center for Theoretical Physics,\\ Universit\"{a}t Bonn, D-53115 Bonn, Germany}
\affiliation{Institute for Advanced Simulation (IAS-4), Forschungszentrum J\"{u}lich, D-52425 J\"{u}lich, Germany}
\affiliation{Peng Huanwu Collaborative Center for Research and Education, International Institute for Interdisciplinary and Frontiers, Beihang University, Beijing 100191, China}

\author{Bernard~Ch.~Metsch\orcidlink{0000-0003-0631-4150}}
\email{metsch@hiskp.uni-bonn.de}
\affiliation{Helmholtz-Institut f\"{u}r Strahlen- und Kernphysik and Bethe Center for Theoretical Physics,\\ Universit\"{a}t Bonn, D-53115 Bonn, Germany}
\affiliation{Institute for Advanced Simulation (IAS-4), Forschungszentrum J\"{u}lich, D-52425 J\"{u}lich, Germany}

\author{Helen~Meyer\orcidlink{0009-0002-2643-0933}}
\email{hmeyer@hiskp.uni-bonn.de}
\affiliation{Helmholtz-Institut f\"{u}r Strahlen- und Kernphysik and Bethe Center for Theoretical Physics,\\ Universit\"{a}t Bonn, D-53115 Bonn, Germany}

\date{\today}

\begin{abstract}
We investigate the fine-tuning of radiative alpha-particle capture on carbon,
$\nuc{12}{C}(\alpha,\gamma)\nuc{16}{O}$, 
at astrophysical energies. Utilizing
results from cluster effective field theory for this reaction, we find that
the low-energy data of the astrophysical S-factor allow for only very small variations in the electromagnetic fine-structure constant $\alpha$, namely
$|\delta \alpha/\alpha| \leq 0.2\,$\textperthousand, in both the $E1$ and the $E2$ radiative capture.
\end{abstract}


\maketitle
\section{Introduction}

Fine-tunings appear in most fields of physics and often pave the
way to new approaches to the pertinent problems, see e.g. Refs.~\cite{Dicke:1961rju,Dvali:1994ms,Arkani-Hamed:2004ymt,Barnes:2011zh,Adams:2019kby,Rende:2024ane,Charalambous:2025ekl}, and for introductory works we
refer to Refs.~\cite{DeVuyst:2020wzv,Sloan:2020zer}. In addition, the 
discussion of fine-tunings is taken up in metaphysics and philosophy~\cite{Friederich:2019kcf,Helbig:2023aro,Goff:2024wvq,Saad:2025rmg},
and for a critical discussion of fine-tuning arguments for extension of the
Standard Model, see~\cite{Grinbaum:2009sk}. The synthesis of the elements shortly after the Big Bang and in stars features a number of fine-tunings. We mention the
so-called deuterium bottleneck in primordial nucleosynthesis at about one minute after the Big Bang, related to the very shallow binding of the lightest nucleus and the multiplicity of sufficiently high-energetic photons ($E_\gamma \geq 2.2\,$MeV) at this time.
Generally, the measured abundances of the light elements, especially $^2$H and
$^4$He, set limits on possible variations of the Higgs vacuum expectation value $v$,
or, equivalently, the light quark masses~\cite{Meyer:2024auq,Meissner:2025jfs}
and the electromagnetic fine-structure constant $\alpha$~\cite{Meissner:2023voo},
see also earlier work in Refs.~\cite{Bergstrom:1999wm,Nollett:2002da,Uzan:2002vq,Coc:2006sx,Dent:2007zu,Iocco:2008va,Bedaque:2010hr,Berengut:2013nh,Pitrou:2018cgg,Burns:2024ods}.
Another much discussed fine-tuning appears in carbon production in hot, old stars,
namely through the famous Hoyle resonance~\cite{Hoyle:1954zz} in the spectrum of $^{12}$C. The Hoyle state is located close to the
$^4$He+$^8$Be threshold thus leading to a resonant enhancement of 
carbon and later oxygen production. 
The triple-alpha reaction rate depends exponentially on the distance of the Hoyle state from this threshold.
The dependence of the resonance condition on the quark masses and the electromagnetic fine-structure constant can be model-independently investigated in
the framework of nuclear lattice effective field theory~\cite{Epelbaum:2012iu,Epelbaum:2013wla,Lahde:2019yvr}. For other works
on this topic and alternative views, see e.g. Refs.~\cite{Livio:1989,Oberhummer:2000zj,Ubaldi:2008nf,Huang:2018kok,Lee:2020tmi,Adams:2022mff}. It should be
noted that the bounds on the light quark masses here are less stringent than the
ones from Big Bang nucleosynthesis (BBN) and also a broader variation in $\alpha$ is allowed. For a nice
discussion of the Hoyle state and its anthropic relevance, see~\cite{Kragh2010-KRAAAM-3}. Differently from this, little is known about the fine-tunings in radiative alpha-particle capture on carbon, 
$\nuc{12}{C}(\alpha,\gamma)\nuc{16}{O}$, 
at astrophysical energies, the so-called holy grail of nuclear astrophysics~\cite{Fowler:1984zz}. 
For a recent review on this reaction we refer to~\cite{deBoer:2025puq}.
Here, we will use results from cluster
effective field theory (EFT)~\cite{Ando:2018lgh,Ando:2025cjk} to work out 
the $\alpha$-dependence of this fundamental reaction and find novel and
more stringent bounds than obtained earlier from element generation in the Big Bang and in stars. Note that within this framework, we cannot address
the issue of a possible time-dependence of the fine-structure constant.

\section{Methodology}
\begin{figure}[tb]
  \begin{center}
   \includegraphics[width=1.0\linewidth, trim=0 0 0 0, clip]{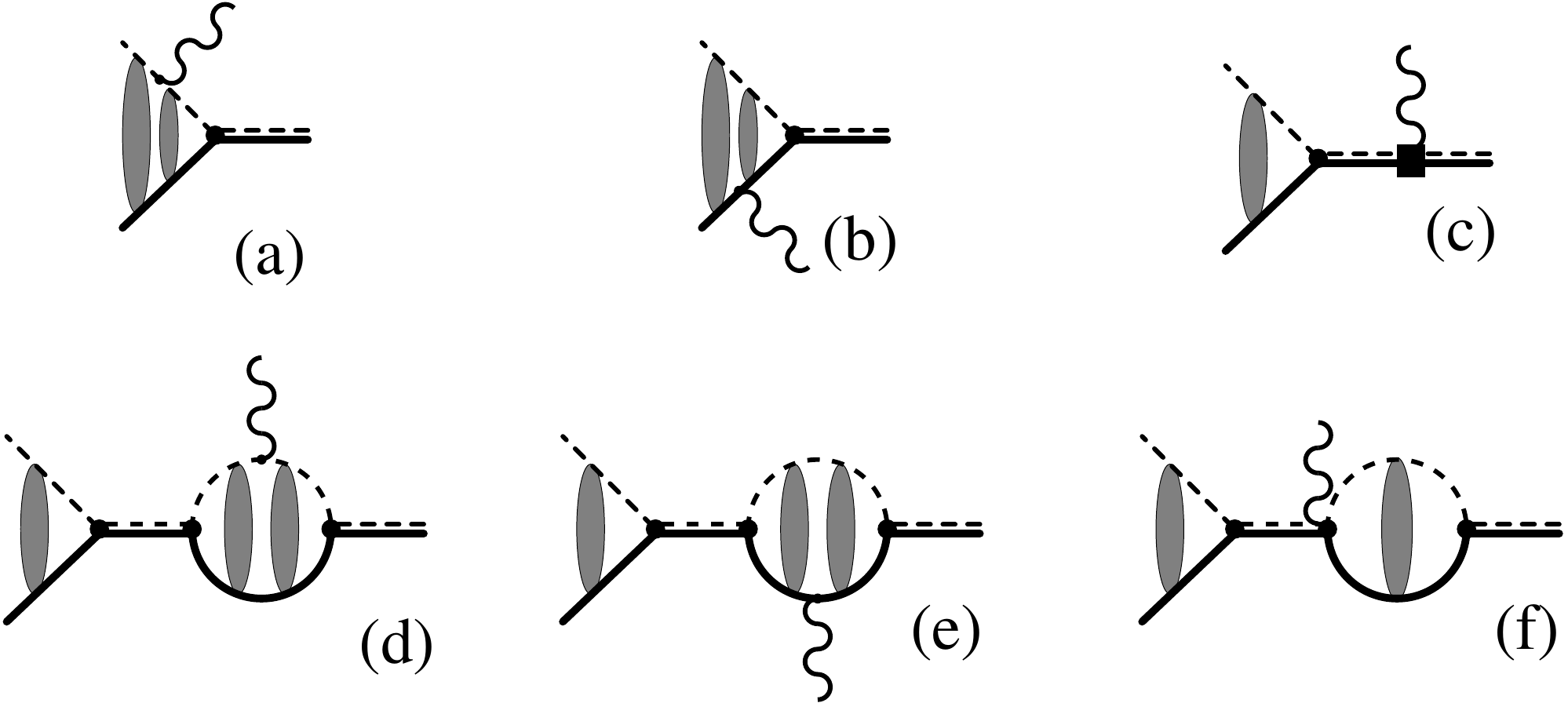}
  \caption{Diagrams of amplitudes for radiative $\alpha$ capture on $\nuc{12}{C}$. A wavy line denotes the outgoing photon, the thin dashed line the $\nuc{4}{He}$ and the solid line the $\nuc{12}{C}$ state. 
  The double thin-dashed / solid lines represent the dressed propagation of the $\nuc{16}{O}$ dimer in the intermediate and final state. 
  The shaded ellipses represent the Coulomb-interaction. 
  Diagrams (a) and (b) are initial state radiation contributions. 
The vertex in diagram (c) is a counter term proportional to $h^{(\ell)}_r$ to renormalize the infinities from the loop diagrams in (d), (e), (f). Figure adopted from~Ref.~\cite{Ando:2025ibj}.}
    \label{fig:diagrammes}
  \end{center}
\end{figure}

In this part we adopt the formulas given in Refs.~\cite{Ando:2018lgh,Ando:2025ibj} for $E1$- and $E2$-radiative capture, respectively, which are based on cluster effective field theory, see~\cite{Ando:2020wtu} and references therein. Here, the cross section for the radiative capture reaction 
\begin{equation}
  \label{eq:rcreac}
  \nuc{4}{He} + \nuc{12}{C} \to 
  \nuc{16}{O}^*(J=\ell) \to \nuc{16}{O}(\textnormal{g.s.}) + \gamma\,,\quad \ell = 1,2
\end{equation}
is given by
\begin{eqnarray}
  \label{eq:rccs1}
  \sigma_{E1}(E)
  &=&
  \frac{4}{3}\,\frac{\alpha\,\mu\,E_\gamma}{p\left(1+\frac{E_\gamma}{m_O}\right)}\,|X^{(1)}|^2\,,
  \\
  \label{eq:rccs2}
  \sigma_{E2}(E)
  &=&
  \frac{4}{3}\,\frac{\alpha\,\mu\,E_\gamma}{p\left(1+\frac{E_\gamma}{m_O}\right)}\,\frac{1}{5}\,|X^{(2)}|^2\,,
\end{eqnarray}
where 
\begin{equation}
  \label{eq:Egamma}
  E_\gamma(p) \approx B_0 + E - \frac{1}{2\,m_O}\,(B_0+E)^2
\end{equation}
is the energy of the photon in the final state
and $B_0$ is the binding energy of the $\nuc{16}{O}$ ground state (g.s.) with respect to the $\alpha-\nuc{12}{C}$ threshold
in the centre-of-mass system (CMS). $E=p^2/(2\,\mu)$ is the (non-relativistic) kinetic energy in the $\alpha-\nuc{12}{C}$ centre-of-mass system with $\mu = m_\alpha\,m_{\nuc{12}{C}}/(m_\alpha+m_{\nuc{12}{C}})$ the reduced mass. 
 Note that we shall ignore the capture through excited bound states of $\nuc{16}{O}$, such as the $0^+_2$-state at $E_\textnormal{ex}(0^+)=6.049\textnormal{ MeV}$, with subsequent further decays. As in~\cite{Ando:2018lgh} such so-called cascade
transitions are considered to be of minor importance.
The masses and charges of the nuclides involved are denoted as $m_\alpha, m_C, m_O$ and $Z_\alpha, Z_C, Z_O$ for $\alpha \equiv \nuc{4}{He}, \nuc{12}{C}$ and $\nuc{16}{O}$, respectively. 
The astrophysical $S$-factor then follows from the cross-sections via
\begin{equation}
    \label{eq:Sfact}
    S(E) = E\,\sigma(E)\,\expo{2\pi\,\eta}\,,
\end{equation}
where the Sommerfeld parameter $\eta:={k_c}/{p}$ with the inverse Bohr radius $k_c := Z_\alpha\,Z_C\,\mu\,\alpha$ is introduced. 
The amplitudes $X^{(\ell)}$ are written as a sum of the contributions
\begin{equation}
  \label{eq:Yl}
  X^{(\ell)} = X^{(\ell)}_{(a+b)} +  X^{(\ell)}_{(c)} +  X^{(\ell)}_{(d+e)} + X^{(\ell)}_{(f)}\,,
\end{equation}
corresponding to the diagrams as given in Fig.~\ref{fig:diagrammes}, also see Ref.~\cite{Ando:2025ibj}.

For $\ell=1$, the initial state radiation amplitude is given by 
\begin{equation}
  \label{eq:X1ab}
    X^{(1)}_{(a+b)}
    = 
     \frac{2\,y^{(0)}\,\expo{\ii\,\sigma_1}}{k_c}
     \intdif{0}{\infty}{x}\,t^{(1)}(x,\eta)\,,
\end{equation}
where the integrand $t^{(1)}(x,\eta)$ is specified in~Appendix~\ref{appA}. Furthermore, 
\begin{eqnarray}
 \label{eq:X1c}
  X^{(1)}_{(c)}
  &=&
    \frac{2}{3}\,y^{(0)}\,A^{(1)}(p)\,W^{(1)}(r_c)\,,
    \\
  \label{eq:X1de}
  X^{(1)}_{(d+e)}
  &=&
    \frac{2}{3}\,y^{(0)}\,A^{(1)}(p)\,U^{(1)}\left(r_c,\eta\right)\,,
  \\
  \label{eq:X1f}
  X^{(1)}_{(f)}
  &=&
    \frac{2}{3}\,y^{(0)}\,A^{(1)}(p)\,V^{(1)}\,,
\end{eqnarray}
where the amplitude 
\begin{equation}
  \label{eq:AmpA1}
  A^{(1)}(p) = \frac{\expo{\ii\,\sigma_1}\,p\,\sqrt{1+\eta^2}\,C_0(\eta)}{D_1(p)}
\end{equation}
with 
\begin{equation}
  \label{eq:C02}
  C_0^2(\eta) = \frac{2\pi\,\eta}{\expo{2\pi\,\eta}-1}\,,
\end{equation}
the normalization of the Coulomb function with $\ell=0$, contains $D_1$, i.e. the inverse of the propagator of the $\alpha-\nuc{12}{C}$ intermediate state in the $\ell=1$ channel.  
This inverse of the propagator in a scattering channel with angular momentum $\ell$ is given by:
\begin{eqnarray}
    \label{eq:Dell}
    D_\ell(p)
    &=&
    \frac{1}{2}\,r_\ell\,\left(\gamma_\ell^2+p^2\right)
    +
    \frac{1}{4}\,P_\ell\,\left(\gamma_\ell^4-p^4\right)
    +
    Q_\ell\,\left(\gamma_\ell^6+p^6\right)
    \nonumber\\
    &&\qquad 
    +
    2\,k_c\,
    \left[
        H_\ell\left(\ii\,\gamma_\ell\right)
        -
        H_\ell\left(p\right)
   \right]\,,
\end{eqnarray}
in terms of the effective range expansion parameters $r_\ell, P_\ell, Q_\ell$ of $\alpha-\nuc{12}{C}$-scattering with relative angular momentum $\ell$.  
Here, the condition, that for bound states, i.e. states below the $\alpha-\nuc{12}{C}$ threshold,
the denominator $D_\ell$ should vanish at $p=\ii\,\gamma_\ell$ was used, where $\gamma_\ell=\sqrt{2\,\mu\,B_\ell}$ is the binding momentum of a state with binding energy $B_\ell$ with respect to the $\alpha-\nuc{12}{C}$ threshold.  
Furthermore, 
\begin{eqnarray}
  \label{eq:HWl}
  H_\ell(p) 
  &=& W_\ell(p)\,H(\eta) = W_\ell(p)\,H\left(\frac{k_c}{p}\right)\,,
   \nonumber\\
    W_\ell(p) 
    &=& 
    \left(\frac{k_c^2}{\ell^2}+p^2\right)\,W_{\ell-1}(p)\,,
    \quad
  W_0(p) = 1\,,
\end{eqnarray}
with
\begin{equation}
\label{eq:Hdef}
  H(\eta)
  :=
  \psi(\ii\,\eta) 
  + 
  \frac{1}{2\,\ii\,\eta} 
  - 
  \log{\left(\ii\,\eta\right)}
\end{equation}
in terms of the di-gamma function $\psi$.

The definitions of the amplitudes $U^{(1)}, V^{(1)}$ and $W^{(1)}$ are given in Appendix~\ref{appA}.
Furthermore, 
\begin{equation}
    \label{eq:sigmaC}
   \sigma_\ell(p) = \textnormal{arg}{\left[\Gamma(\ell+1+\ii\,\eta)\right]}  
  \,\Leftrightarrow\,
  \expo{\ii\,\sigma_\ell(p)}
  =
  \sqrt{\frac{\Gamma(\ell+1+\ii\,\eta)}{\Gamma(\ell+1-\ii\,\eta)}}
\end{equation}
is the Coulomb phase and the normalization $y^{(\ell)}$ is related to the asymptotic normalization constant $|C_b|_\ell$ via
\begin{equation}
\label{eq:ANC}
    |C_b|_\ell 
    = 
    \frac{(\gamma_\ell)^\ell}{\mu^{\ell-1}}\,
\frac{\Gamma\left(\ell + 1 + \frac{k_c}{\gamma_\ell}\right)}{\Gamma(\ell+1)}\,
\frac{1}{\sqrt{(2\,\ell+1)\,\pi}}\,y^{(\ell)}\,,
\end{equation}
where $\Gamma(x)$ is the gamma function.

Likewise, for $\ell=2$, 
the initial state radiation amplitude is given by 
\begin{equation}
  \label{eq:X2ab}
    X^{(2)}_{(a+b)}
    = 
     -\frac{6\,y^{(0)}\,\expo{\ii\,\sigma_2}}{k_c}
     \intdif{0}{\infty}{x}\,t^{(2)}(x,\eta)\,,
\end{equation}
where the integrand $t^{(2)}(x,\eta)$ is specified in Appendix~\ref{appA}. Furthermore, 
\begin{eqnarray}
 \label{eq:X2c}
  X^{(2)}_{(c)}
  &=&
    \frac{1}{5}\,y^{(0)}\,A^{(2)}(p)\,W^{(2)}(r_c)\,,
    \\
  \label{eq:X2de}
  X^{(2)}_{(d+e)}
  &=&
    \frac{1}{5}\,y^{(0)}\,A^{(2)}(p)\,U^{(2)}\left(r_c,\eta\right)\,,
  \\
  \label{eq:X2f}
  X^{(2)}_{(f)}
  &=&
    \frac{1}{5}\,y^{(0)}\,A^{(2)}(p)\,V^{(2)}\,,
\end{eqnarray}
where the amplitude 
\begin{equation}
  \label{eq:AmpA2}
  A^{(2)}(p) = \frac{\expo{\ii\,\sigma_2}\,p^2\,\sqrt{(1+\eta^2)\,(4+\eta^2)}\,C_0(\eta)}{D_2(p)}
\end{equation}
contains $D_2$, see Eq.~(\ref{eq:Dell}). The amplitudes $U^{(2)}, V^{(2)}$ and $W^{(2)}$ are given in Appendix~\ref{appA}.

The parameters that enter the formulas above are the effective range parameters $r_\ell, P_\ell, Q_\ell$ that determine the inverse of the propagator $D_\ell$. Furthermore, the normalization of the ground state, $y^{(0)}$ (or $|C_b|_0$, see Eq.~(\ref{eq:ANC})) determines the overall magnitude of the amplitude $X^{(\ell)}$, while the renormalized coupling $h^{(\ell)}_c$, depending on the choice of the cut-off $r_c$ in the integrals of Eqs.~(\ref{eq:ampU1},\ref{eq:ampU2}) in Appendix~\ref{appA}, can be tuned in order to account for the observed astrophysical $S$-factor. We finally note that the fine-structure constant $\alpha$ is ubiquitous: The inverse Bohr radius $k_c=Z_\alpha\,Z_C\,\mu\,\alpha$ linearly depends on $\alpha$ and enters, via $\eta(p) = k_c/p, \eta_\ell = k_c/\gamma_\ell$, the parameters and arguments of the Coulomb- and Whittaker functions and, in particular the last term in the inverse propagator $D_\ell$, see Eq.~(\ref{eq:Dell}). 
Note that the nuclear masses, and thus e.g. also $\mu$, depend on $\alpha$, the main effect being the dependence via the Coulomb repulsion  $V_C$ of the protons. This, however, is a relatively small contribution to the nuclear mass ($V_C/m < 0.1\%$ for the nuclei considered here) and therefore the effect from a small variation of $\alpha$ on the nuclear masses can be safely ignored here. 
Note that this also applies to the excitation energies of bound states, parameterized by the binding momenta $\gamma_\ell$. In the present study the position of the subthreshold states with $E_{\textnormal{ex}}(1^-)= 7.117\textnormal{ MeV}$ and $E_{\textnormal{ex}}(2^+)= 6.917\textnormal{ MeV}$ thus remain unaltered when varying $\alpha$.

We concentrate on the dependence of the $S$-factor on the values of the electromagnetic fine-structure constant and consider fractional changes $\delta$:
\begin{equation}
    \label{eq:avar}
    \alpha = \alpha_0\,(1 + \delta)\,,
\end{equation}
where 
\begin{eqnarray}
\label{eq:alpha0}
\alpha_0 &=& 7.297 352 5693(11)\times10^{-3}\nonumber\\ &=& 1/137.035 999 084(21)
\end{eqnarray}
is the nominal value from the PDG~\cite{ParticleDataGroup:2024cfk}, and we assume $|\delta| \ll 0.1$.

The initial state radiation amplitudes $X^{(\ell)}_{a+b}$ are of minor significance for the resulting cross-sections as already observed in Ref.~\cite{Ando:2018lgh}. The amplitudes $X^{\ell}_{(c)}$ and 
$X^{\ell}_{(f)}$ interfere constructively and their sum cancels to a large extent the contribution from $X^{\ell}_{(d+e)}$. Within the present framework the radiative capture cross-section is thus extremely fine-tuned: By a judicious choice of the renormalized couplings $h^{(\ell)}_r$, which enters the amplitudes $X^{\ell}_{(c)}$, the resulting cross-sections (or S-factors) can be tuned to the experimental values. Its energy dependence is then found to be dominated by the inverse of the propagator $D_\ell(p)$, see Eq.~(\ref{eq:Dell}), in the amplitudes $A^{(\ell)}(p)$. 

At this point we like to stress that we do not aim to determine all the parameters in order to achieve the best description of the $S$-factors within the present framework and therefore refrain from a detailed error analysis. The purpose here is merely to explore how, given a reasonable description of the experimental data, a change in the fine-structure constant $\alpha$ would affect the results.

\FloatBarrier
\section{Results and discussion}

\begin{figure}[!htb]
  \begin{center}
   \includegraphics[width=1.0\linewidth, trim=50 220 50 70, clip]{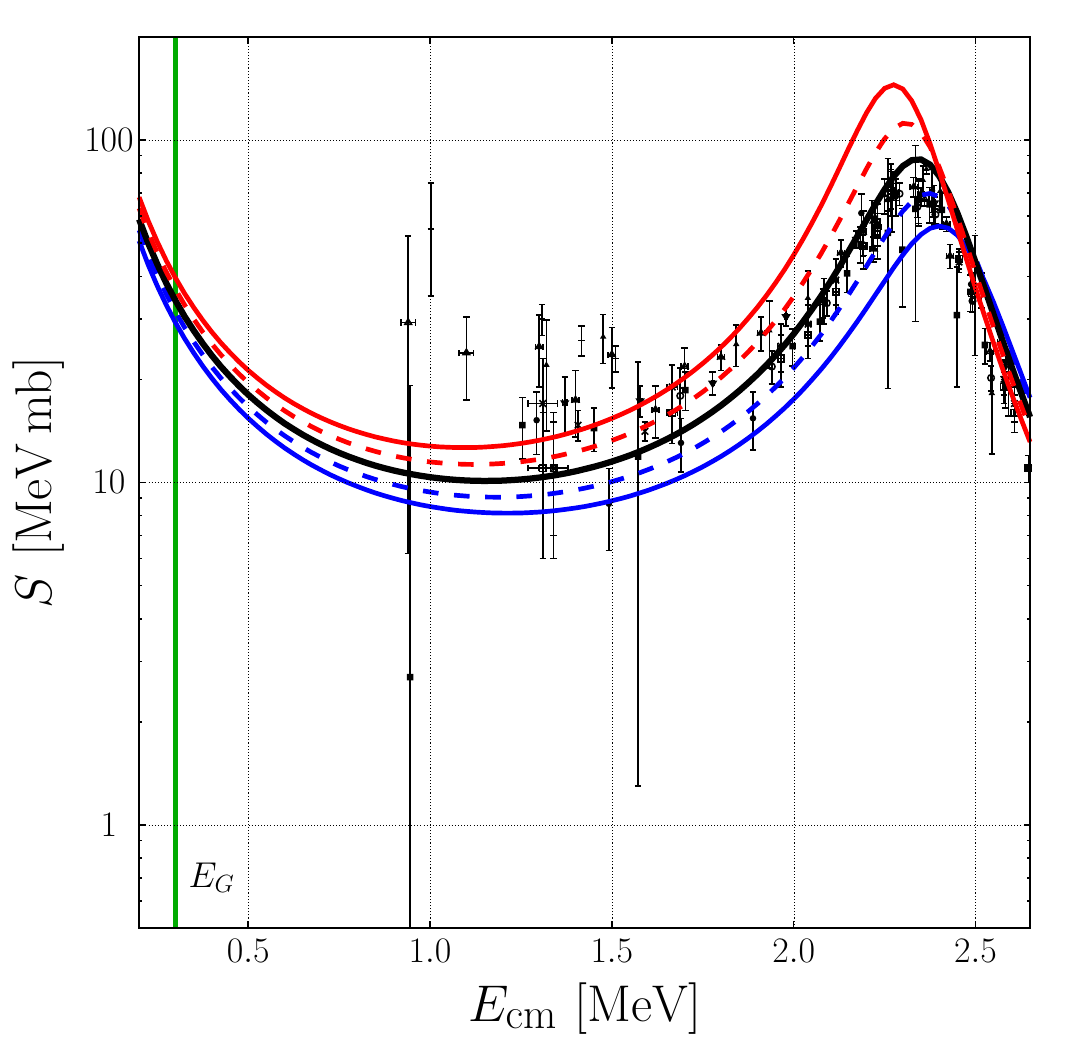}
  \caption{Fine-structure constant ($\alpha = (1+\delta)\,\alpha_0)$ variation of the astrophysical $S$-factor of the $\nuc{4}{He}+\nuc{12}{C} \to \nuc{16}{O}(1^-) \to \nuc{16}{O}(0^+)$ radiative capture. The result for the nominal value $\alpha_0$ is displayed in black. The blue (dashed) curves correspond to $\delta = -0.0002 (-0.0001)$; the red (dashed) curves to $\delta=0.0002 (0.0001)$\,. 
  The position of the Gamow energy at $E_G \simeq 0.3\,\unit{MeV}$ is indicated by a vertical green line. 
  The data are from Refs.~\cite{Plag:2012zz,Makii:2009zz,Assuncao:2006vy,Kunz:2001zz,Roters:1999xyz,Kremer:1988zz,Redder:1987xba,Gialanella:2001,Ouellet:1996zz}.}
    \label{fig:SalphavarE1001}
  \end{center}
\end{figure}

We first considered different normalizations for $\ell=1,2$ and found that the $S_{E1}$-factor is best described by a large value of the normalization $y^{(0)}$, whereas the $S_{E2}$-factor profits from a rather low value for this normalization. In the present study for consistency we opted for the universal value $y^{(0)}=0.0798\,\unit{MeV}^{-\frac{1}{2}}$
 (corresponding to the asymptotic normalization $|C_b|_0=10.0\,\unit{fm}^{-\frac{1}{2}}$) as a compromise.  
As values for the effective range parameters for the $E1$-radiative capture we used 
the parameters from Ref.~\cite{Ando:2022flx} : 
$r_1=0.415314\, \unit{fm}^{-1}$, 
$P_1=-0.57427\,\unit{fm}$ 
and 
$Q_1 = 0.02032\,\unit{fm}^3$. 
For $r_c=0.01\,\unit{fm}$ with the effective coupling $h^{(1)}_r=272125\,\unit{MeV}^{3}$ a reasonable description of the $S$-factor was found. The resulting $S$-factor is displayed with the nominal value $\alpha_0$ of Eq.~\eqref{eq:alpha0} as the black solid line in Fig.~\ref{fig:SalphavarE1001}. 
It is well known, that isoscalar $E1$-transitions are suppressed, see e.g. the discussion in Ref.~\cite{Ando:2018lgh} and references therein.
The dependence on the 
cut-off $r_c$ is weak and discussed in Appendix~\ref{appB}.

Note that for $\ell=1$ the effective range parameters are such that 
the real part of the inverse propagator $D_1(p(E))$ vanishes at $E = 2.438\,\unit{MeV}$ and thus accounts for the $1^-$ resonance ($E_x=9.585(11)\,\unit{MeV}, \Gamma = 0.42(2)\,\unit{MeV}$), corresponding to a CMS-energy $E= 2.42\,\unit{MeV}$ of $\nuc{16}{O}$ also observed in the radiative capture studied here. 

In a similar fashion the $S$-factor for $E2$-radiative capture was calculated with the effective range parameters 
$r_2=0.157453\, \unit{fm}^{-3}$, 
$P_2=-1.04781\,\unit{fm}^{-1}$ 
and 
$Q_2 = 0.1403\,\unit{fm}$,  
taken from Ref.~\cite{Ando:2025ibj}. Note that in this reference a slightly smaller value for the ground state normalization 
$y^{(0)}=0.058\,\unit{MeV}^{-\frac{1}{2}}$
(corresponding to $|C_b|_0 = 7.27\,\unit{fm}^{-\frac{1}{2}}$)
was used. 
With $r_c=0.01\,\unit{fm}$ the optimal value of the 
effective coupling was found to be  
$h^{(2)}_r = 4.591\,\times\,10^{12}\,\unit{MeV}^4$.
The result 
with the nominal value $\alpha_0$ of Eq.~\eqref{eq:alpha0} 
is represented as the black solid line in 
Fig.~\ref{fig:SalphavarE2001}. The dependence on the 
cut-off $r_c$ is again weak and is also discussed in Appendix~\ref{appB}.
\begin{figure}[!htb]
  \begin{center}
   \includegraphics[width=1.0\linewidth, trim=50 220 50 70, clip]{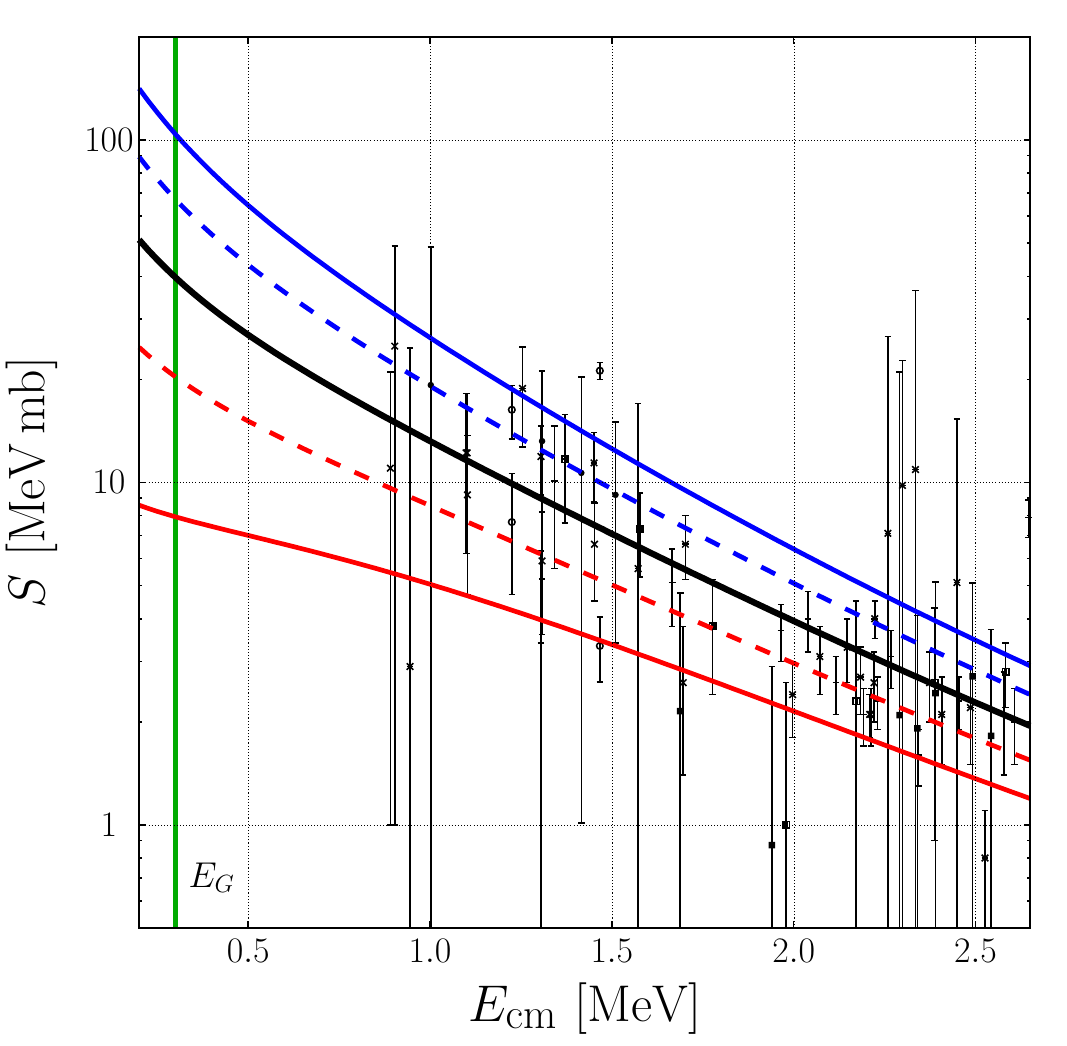}
  \caption{Fine-structure constant ($\alpha = (1+\delta)\,\alpha_0)$ variation of the astrophysical $S$-factor of the $\nuc{4}{He}+\nuc{12}{C} \to \nuc{16}{O}(2^+) \to \nuc{16}{O}(0^+)$ radiative capture. The result for the nominal value $\alpha_0$ is displayed in black. The blue (dashed) curves correspond to $\delta = -0.0002 (-0.0001)$; the red (dashed) curves to $\delta=0.0002 (0.0001)$\,.
  The position of the Gamow energy at $E_G = 0.3\,\unit{MeV}$ is indicated by a vertical green line. The data are from Refs.~\cite{Plag:2012zz,Makii:2009zz,Kunz:2001zz,Roters:1999xyz,Ouellet:1996zz,Fey:2004}.
  }
    \label{fig:SalphavarE2001}
  \end{center}
\end{figure}

For $\ell=2$ the effective range parameters do not describe the narrow $2^+$ resonance of $\nuc{16}{O}$ at $E_x = 9.85$\,\unit{MeV} ($\Gamma = 0.624\,\unit{keV}$); corresponding to a CMS-energy $E = 2.68\,\unit{MeV}$, just outside the energy range displayed in Fig.~\ref{fig:SalphavarE2001}. In the analysis of the elastic phase shifts of Ref.~\cite{Ando:2025ibj} it was explicitly introduced as a resonant contribution.  

\begin{figure}[!htb]
    \centering
    \includegraphics[width=0.7\linewidth, trim=50 220 50 70, clip]{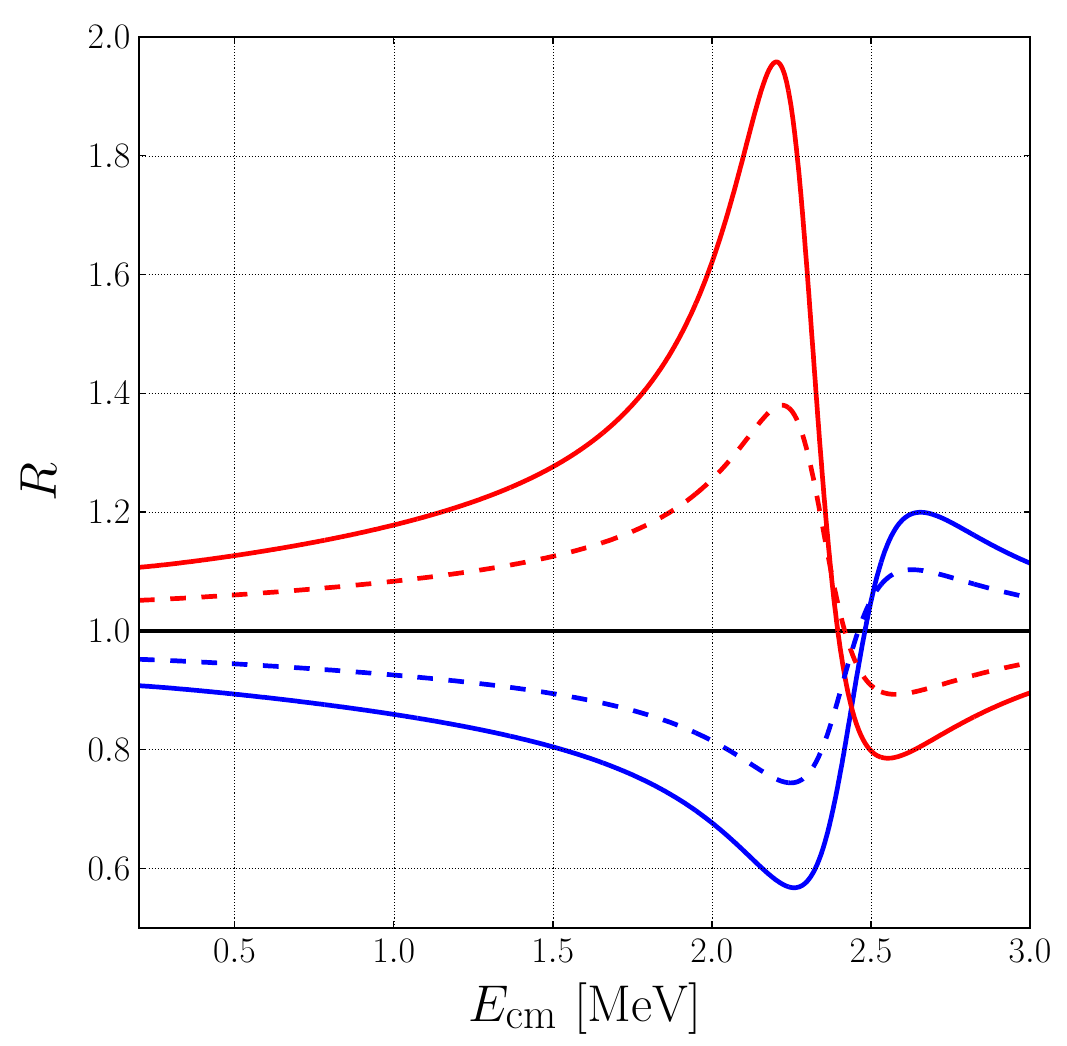}
    \caption{Ratio of the $S$-factor for radiative $E1$-capture accounting only for the change in the amplitude $A^{(1)}$ of Eq.~(\ref{eq:AmpA1}).
    The red (dashed) curve is the ratio for $\delta=0.0002 (0.0001)$, the blue (dashed) curve for $\delta=-0.0002 (-0.0001)$.}
    \label{fig:ratio}
\end{figure}

Concerning $E1$-radiative capture, a variation of $\alpha$ according to Eq.~(\ref{eq:avar}) with $\delta \in [-0.0002,0.0002]$ leads to an appreciable change in the $S$-factor: In particular, as displayed in Fig.~\ref{fig:SalphavarE1001}, the position of the $1^-$-resonance peak shifts by 
$\Delta(E_R) = 0.06\,\unit{MeV}$, while the peak value changes by almost a factor 2 for a variation in this range. This is almost entirely due to the corresponding change in the amplitude $A^{(1)}$ of Eq.~(\ref{eq:AmpA1}) with such a variation. Indeed, assuming that, when varying $\alpha$ the ratio of the cross-section is simply given by
\begin{equation}
\label{eq:E1ratio}
    r = \frac{|A^{(1)}(\alpha)|^2}{|A^{(1)}(\alpha_0)|^2}
\end{equation}
the main features observed in Fig.~\ref{fig:SalphavarE1001} are accounted for, as is illustrated in Fig.~\ref{fig:ratio}, where the ratio $R$ of the corresponding $S$-factor is plotted. The sum of the amplitudes $|U^{(1)}+V^{(1)}+W^{(1)}|$ varies 
by less than $3\,\%$ 
when varying $\alpha$ in the range discussed here and this variation is thus of minor importance.  

No such single dominant effect could be identified for the $\alpha$ variation of the $E2$-radiative capture, but as shown in Fig.~\ref{fig:SalphavarE2001}, the sensitivity was found to be of the same order of magnitude. In contrast to the variation of the electric dipole radiative capture the variation of the amplitude $|A^{(2)}|$ for electric quadrupole radiative capture is only $\approx 2\,\%$, but here, although individually $|U^{(2)}|$, $|V^{(2)}|$ and $|W^{(2)}|$ vary by at most 
$\approx 3\,\%$, due to the interference discussed above the sum of the amplitudes $|U^{(2)}+V^{(2)}+W^{(2)}|$ varies by as much as 
$\approx 55\,\%$ and $|A^{(2)}\,(U^{(2)}+V^{(2)}+W^{(2)})|$ by 
$\approx 60\,\%$, leading to the 
change in the $S$-factor observed in Fig.~\ref{fig:SalphavarE2001}. Finally note that the variation of $\alpha$ considered here changes the factor in the conversion of the cross-section to the  astrophysical $S$-factor, via the occurrence of the Sommerfeld parameter $\eta=k_c/p$ in Eq.~(\ref{eq:Sfact}) by less than $\approx 1\,\%$ for energies above the Gamow-energy $E_G$ and this is thus a minor effect. 

\section{Summary and outlook}

In the framework of cluster effective field theory, we have investigated the dependence of the astrophysical S-factor of the radiative 
alpha-particle capture on carbon at astrophysical energies on the fine-structure constant $\alpha$. 
We find that for both $E1$ and $E2$ radiative transitions, such variations
cannot exceed $|\delta \alpha/\alpha| \leq 0.2\,$\textperthousand ~by comparison with the data. While in the $E1$ transition this is due to
the $\alpha$-dependence of the inverse propagator in the $\alpha-^{12}{\rm C}$ channel, for the $E2$ transition, no such single dominant effect could be
identified. Rather in this case it is the strong cancellation of the various amplitudes with the photon in the final state that generates
this strong bound. This bound is much stronger than found for other element generations in stars, such as the Hoyle state in $^{12}$C, see
e.g.~\cite{Lahde:2019yvr}, or from element generation in the Big Bang~\cite{Meissner:2023voo}. It would be interesting to confirm this bound
using nuclear lattice effective field theory along the lines for $\alpha$-$\alpha$ scattering~\cite{Elhatisari:2021eyg}, but such a calculation
requires sizeable computational resources and is unlikely to be done in the near future. Finally, we note that the quark mass dependence of the radiative $\alpha$ capture on carbon is more difficult to assess, either by calculating the quark mass dependence of the pertinent scattering parameters or directly within Nuclear Lattice EFT (NLEFT, for a detailed introduction and overview see \cite{Lahde:2019npb}) .

\section*{Acknowledgements}

This work was supported in part by the European
Research Council (ERC) under the European Union's Horizon 2020 research
and innovation programme (grant agreement No. 101018170),
and by the CAS President's International Fellowship Initiative (PIFI) (Grant No.~2025PD0022).  The authors gratefully acknowledge the Gauss Centre for Supercomputing e.V. (www.gauss-centre.eu)
for funding this project by providing computing time on the GCS Supercomputer JUWELS
at J\"ulich Supercomputing Centre (JSC) and the support of the project \mbox{EXOTIC} by the JSC by dedicated HPC time provided on the \mbox{JURECA DC} GPU partition. Furthermore, the authors gratefully acknowledge the computing time provided on the high-performance computer
HoreKa by the National High-Performance Computing Center at KIT (NHR@KIT). This center is
jointly supported by the Federal Ministry of Education and Research and the Ministry of Science,
Research and the Arts of Baden-Württemberg, as part of the National High-Performance Computing
(NHR) joint funding program (https://www.nhr-verein.de/en/our-partners). HoreKa is partly funded
by the German Research Foundation (DFG).

%


\appendix
\label{sec:app}

\begin{widetext}


\section{Expressions for the radiative capture amplitudes}
\label{appA}

Below we specify the formulas for the radiative capture amplitudes. These were adapted from Refs.~\cite{Ando:2018lgh,Ando:2025ibj}.

The amplitudes for $\ell=1$ radiative capture, with $\eta:={k_c}/{p}, x := k_c\,r$ are given by 
\begin{equation}
\label{eq:ampT1}
  t^{(1)}(x,\eta)
  :=
    x\,\Gamma(1+\eta_0)\,
    W_{-\eta_0,\frac{1}{2}}\left(\frac{2\,x}{\eta_0}\right)
    \left[
        \frac{Z_\alpha\,\mu}{m_\alpha}\,j_0\left(\frac{\mu}{m_\alpha}\,\frac{x}{\eta_\gamma(\eta)}\right)
       -
       \frac{Z_C\,\mu}{m_C}\,j_0\left(\frac{\mu}{m_C}\,\frac{x}{\eta_\gamma(\eta)}\right)
    \right]
    \left\lbrace
        \frac{1}{x}\,F_1'\left(\eta,\frac{x}{\eta}\right)
        +
        \frac{\eta}{x^2}\,F_1\left(\eta,\frac{x}{\eta}\right)
    \right\rbrace~,
\end{equation}
with $F_\ell(\eta,z)$ representing the regular Coulomb function, $F_\ell'(\eta,z)=\diff{}{z}\,F_{\ell}(\eta,z)$ its derivative and $j_\ell$ the regular spherical Bessel function. 
With the dimensionless quantities $\eta_0 := k_c/\gamma_0$, $\eta_\gamma(\eta) := k_c/E_\gamma(p)$ we also define the amplitudes:
\begin{eqnarray}
  \label{eq:ampU1}
  u^{(1)}(x,\eta)
  &:=&
       x\,\Gamma(1+\eta_0)\,\Gamma(2+\ii\,\eta)\,
       W_{-\eta_0,\frac{1}{2}}\left(\frac{2\,x}{\eta_0}\right)
       \left[
       \frac{Z_\alpha\,\mu}{m_\alpha}\,j_0\left(\frac{\mu}{m_\alpha}\,\frac{x}{\eta_\gamma(\eta)}\right)
       -
       \frac{Z_C\,\mu}{m_C}\,j_0\left(\frac{\mu}{m_C}\,\frac{x}{\eta_\gamma(\eta)}\right)
       \right]
  \nonumber\\
  &&
     \quad\times
     \left\lbrace
     -
     \frac{2\,\ii}{x\,\eta}\,W_{-\ii\,\eta,\frac{3}{2}}'\left(-2\,\ii\,\frac{x}{\eta}\right)
     +
     \frac{1}{x^2}\,W_{-\ii\,\eta,\frac{3}{2}}\left(-2\,\ii\,\frac{x}{\eta}\right)
     \right\rbrace\,,
     \nonumber\\
  U^{(1)}(r_c,\eta)
  &:=&
       \frac{\ii\,k_c}{\eta}\intdif{k_c r_c}{\infty}{x}\,u^{(1)}(x,\eta)\,,
\end{eqnarray}
where $
W_{\kappa,\mu}'(z)
=
\diff{}{z}\,W_{\kappa,\mu}(z)
$ is the derivative of the Whittaker function $W_{\kappa,\mu}$\,, 
as well as the constant
\begin{eqnarray}
  \label{eq:ampV1}
  V^{(1)}
  &:=& -\frac{9}{2}
       \left[
       \frac{Z_\alpha\,\mu}{m_\alpha}
       -
       \frac{Z_C\,\mu}{m_C}
       \right]
       \left\lbrace
       -2\,k_c\,H(-\ii\,\eta_0)
       \right\rbrace\,.
\end{eqnarray}
Finally, we define the function
\begin{eqnarray}  
\label{eq:ampW1}
  W^{(1)}(r_c)
  &:=& \frac{9\pi\,Z_O}{\mu\,m_O}
       \left\lbrace
       h^{(1)}_r +
       \frac{k_c\,\mu}{9\pi}\,\frac{m_O}{Z_O}
       \left[
       \frac{Z_\alpha\,\mu}{m_\alpha}
       -
       \frac{Z_C\,\mu}{m_C}
       \right]
       \left(
       \log{\left(\frac{\lambda}{2}\,r_c\right)}
       -
       9\,
       \log{\left(\frac{\lambda}{2\,k_c}\right)}
       \right)
       \right\rbrace\,,
\end{eqnarray}
where $\lambda$ is the scale of dimensional regularization, chosen as in Ref.~\cite{Ando:2025cjk} as the high-momentum scale $\lambda=\Lambda_H = 160\,\unit{MeV}$.  

The amplitudes for $\ell=2$ radiative capture have a similar structure:
\begin{equation}
\label{eq:ampT2}
  t^{(2)}(x,\eta)
  :=
    x\,\Gamma(1+\eta_0)\,
    W_{-\eta_0,\frac{1}{2}}\left(\frac{2\,x}{\eta_0}\right)
    \left[
        \frac{Z_\alpha\,\mu}{m_\alpha}\,j_1\left(\frac{\mu}{m_\alpha}\,\frac{x}{\eta_\gamma(\eta)}\right)
       +
       \frac{Z_C\,\mu}{m_C}\,j_1\left(\frac{\mu}{m_C}\,\frac{x}{\eta_\gamma(\eta)}\right)
    \right]
    \left\lbrace
        \frac{1}{x}\,F_2'\left(\eta,\frac{x}{\eta}\right)
        +
        \frac{2\,\eta}{x^2}\,F_2\left(\eta,\frac{x}{\eta}\right)
    \right\rbrace\,.
\end{equation}
Also
\begin{eqnarray}
  \label{eq:ampU2}
  u^{(2)}(x,\eta)
  &:=&
       x\,\Gamma(1+\eta_0)\,\Gamma(3+\ii\,\eta)\,
       W_{-\eta_0,\frac{1}{2}}\left(\frac{2\,x}{\eta_0}\right)
       \left[
       \frac{Z_\alpha\,\mu}{m_\alpha}\,j_1\left(\frac{\mu}{m_\alpha}\,\frac{x}{\eta_\gamma(\eta)}\right)
       +
       \frac{Z_C\,\mu}{m_C}\,j_1\left(\frac{\mu}{m_C}\,\frac{x}{\eta_\gamma(\eta)}\right)
       \right]
  \nonumber\\
  &&
     \quad \times
     \left\lbrace
     -
     \frac{2\,\ii}{x\,\eta}\,W_{-\ii\,\eta,\frac{5}{2}}'\left(-2\,\ii\,\frac{x}{\eta}\right)
     +
     \frac{2}{x^2}\,W_{-\ii\,\eta,\frac{5}{2}}\left(-2\,\ii\,\frac{x}{\eta}\right)
     \right\rbrace\,,
     \nonumber\\
  U^{(2)}(r_c,\eta)
  &:=&
       \frac{k_c^2}{\eta^2}\,\intdif{k_c r_c}{\infty}{x}\,u^{(2)}(x,\eta)\,,
\end{eqnarray}
as well as the constant
\begin{eqnarray}
  \label{eq:ampV2}
  V^{(2)}
  &:=& -\frac{75}{4}\,k_\gamma\,
       \left[
       \frac{Z_\alpha\,\mu^2}{m_\alpha^2}
       +
       \frac{Z_C\,\mu^2}{m_C^2}
       \right]
       \left\lbrace
       -2\,k_c\,H(-\ii\,\eta_0)
       \right\rbrace\,,
\end{eqnarray}
and
\begin{eqnarray}  
\label{eq:ampW2}
  W^{(2)}(r_c)
  &:=& \frac{25\pi\,Z_O}{\mu\,m_O^2}\,k_\gamma\,
       \left\lbrace
       -h^{(2)}_r +
       \frac{3\,k_c\,\mu}{2\pi}\,\frac{m_O^2}{Z_O}
       \left[
       \frac{Z_\alpha\,\mu^2}{m_\alpha^2}
       +
       \frac{Z_C\,\mu^2}{m_C^2}
       \right]
       \left(
       \frac{4}{225}\,\log{\left(\frac{\lambda}{2}\,r_c\right)}
       -
    \log{\left(\frac{\lambda}{k_c}\right)} 
       \right)
       \right\rbrace\,.
\end{eqnarray}

\section{Results with other cut-off parameters}
\label{appB}
The resulting cross sections depend on the value $r_c$ of the cut-off used to regularize the integrals of Eqs.~(\ref{eq:ampU1},\ref{eq:ampU2}). Here we briefly discuss this cut-off dependence. For the values of the cut-off $r_c=0.01\,\unit{fm}, 0.05\,\unit{fm}, 0.10\,\unit{fm}$
an optimal value of the renormalized coupling $h^{(\ell)}_r$ was chosen in order to account for the experimental value of the resulting $S$-factor at $E \approx 2.2\, \unit{MeV}$. Since the cross-section depends quadratically on the amplitudes in fact two values of $h^{(\ell)}_r$ are found in this manner; these two values (labeled "low" and "high") are listed in Table~\ref{tab:parE1} and Table~\ref{tab:parE2} for $E1$- and $E2$-radiative capture, respectively. 
description

\renewcommand{\arraystretch}{1.2}
\begin{table}[htp]
\caption{
Parameters entering the calculation of $E1$-radiative capture: 
The values of the effective range parameters are : 
$r_1=0.415314\, \unit{fm}^{-1}$, 
$P_1=-0.57427\,\unit{fm}$, 
$Q_1 = 0.02032\,\unit{fm}^3$, see Ref.~\cite{Ando:2022flx}.
The normalization $y^{(0)}=0.0798\,\unit{MeV}^{-\frac{1}{2}}$
corresponds to $|C_b|_0=10.0\,\unit{fm}^{-\frac{1}{2}}$.
For each choice of the cut-off $r_c$ the 
effective coupling $h^{(1)}_r$ in this work was chosen such as 
to obtain an optimal description of the $S$-factor. 
}
\label{tab:parE1}
\sisetup{
  round-mode=places,
  scientific-notation = fixed,
}

\begin{tabular}{
  S[round-precision=2, 
  table-format=2.2]
  S[round-precision=3, 
  fixed-exponent = 5, 
  minimum-decimal-digits = 3, 
  table-format=4.3e1]
  S[round-precision=3, 
  fixed-exponent = 5, 
  minimum-decimal-digits = 3, 
  table-format=4.3e1]
}
\toprule
{$r_c\,[\unit{fm}]$} & \multicolumn{2}{c}{$h^{(1)}_r\,[\unit{MeV}^3]$} \\  
\cline{2-3}
& "low" & "high" \\
\midrule
0.01 & 268750 & 277000 \\
0.05 & 284000 & 292500 \\ 
0.10 & 297250 & 305500 \\
\bottomrule
\end{tabular}

\end{table}

In fact, for $E1$-radiative capture the best decription of the energy dependence of the $S$-factor was found for the "low" values, the "high" values leading to results that are much too low at energies $E < 2\,\unit{MeV}$. The results are displayed in Fig.~\ref{fig:SalphavarE1001} for $r_c=0.01\,\unit{fm}$ and in Fig.~\ref{fig:SalphavarE1005010} for $r_c=0.05, 0.10\,\unit{fm}$. 
\begin{figure}[!htb]
  \centering
  \includegraphics[width=0.4\textwidth, trim=50 220 50 70, clip]{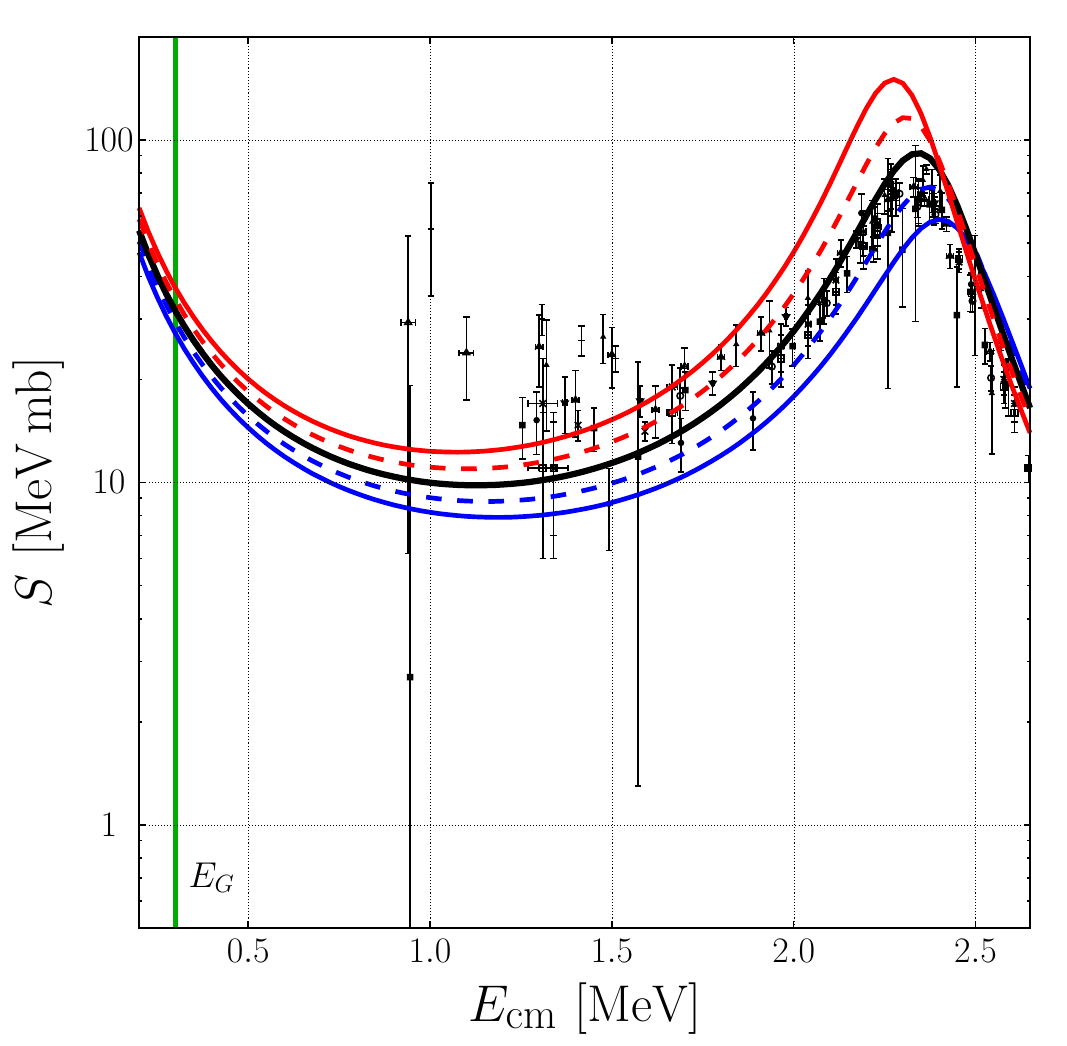}\hspace*{3em}
  \includegraphics[width=0.4\textwidth, trim=50 220 50 70, clip]{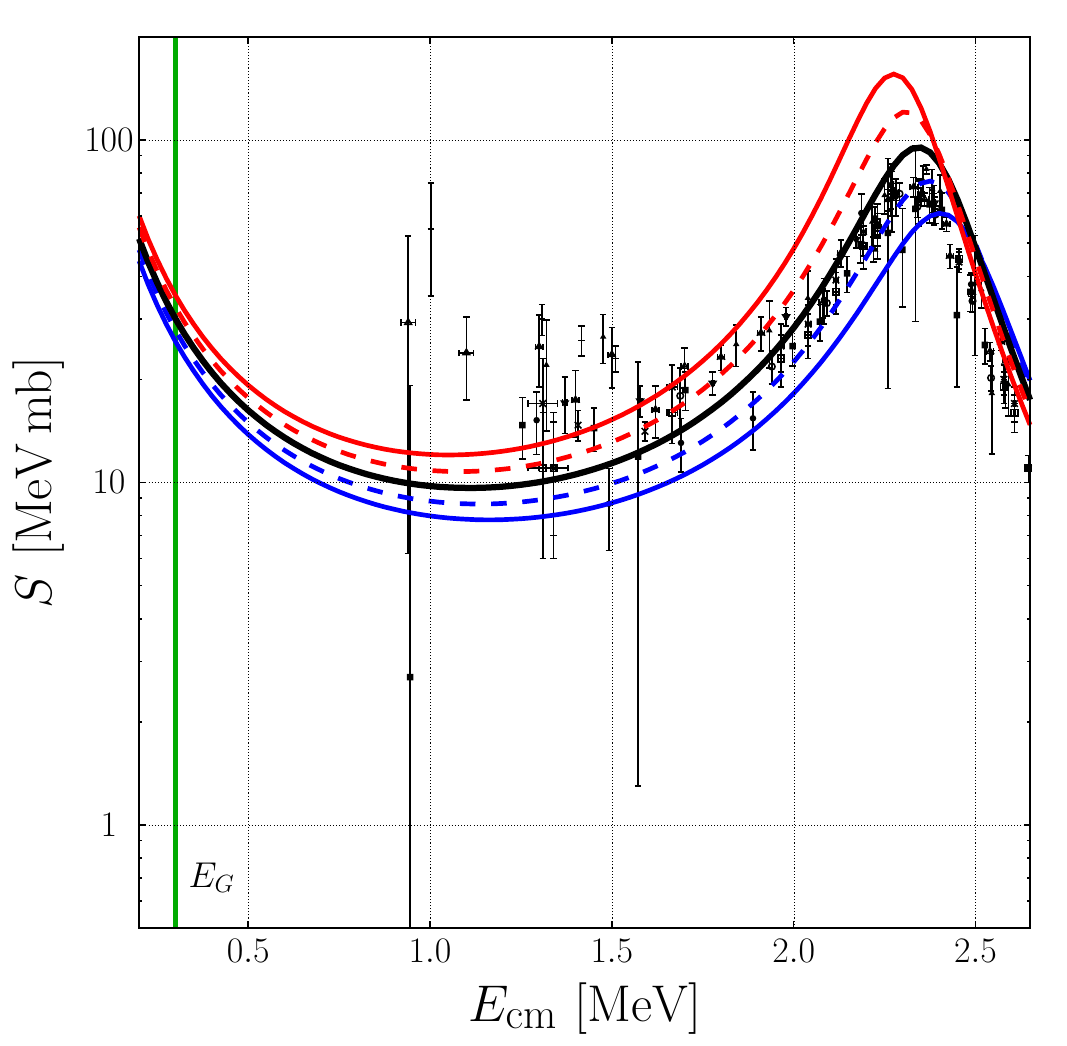}
  \caption{
    Fine structure constant ($\alpha = (1+\delta)\,\alpha_0)$
    variation of the astrophysical $S$-factor of the
    $\nuc{4}{He}+\nuc{12}{C} \to \nuc{16}{O}(1^-) \to \nuc{16}{O}(0^+)$
    radiative capture for two alternative values of the coordinate space cutoff:
    $r_c=0.05\,\unit{fm}$ (left), $r_c=0.10\,\unit{fm}$ (right).
For the parameters, see Table~\ref{tab:parE1}. Also see the caption to Fig.~\ref{fig:SalphavarE1001}.}
 \label{fig:SalphavarE1005010} 
\end{figure} 

In contrast to the electric dipole case the best values for $E2$-radiative capture were found with the "high" values for $h^{(2)}_r$; with the low values the $S$-factor was found to be much too large for energies $E < 2\,\unit{MeV}$. The results are presented in Fig.~\ref{fig:SalphavarE2001} for $r_c=0.01\,\unit{fm}$ and in Fig.~\ref{fig:SalphavarE2005010} for $r_c = 0.05, 0.10\,\unit{fm}$.

\begin{table}[htp]
\caption{
Parameters entering the calculation of $E2$-radiative capture: 
The values of the effective range parameters 
$r_2=0.157453\, \unit{fm}^{-3}$, 
$P_2=-1.04781\,\unit{fm}^{-1}$ 
and 
$Q_2 = 0.1403\,\unit{fm}$ were taken from Ref.~\cite{Ando:2025ibj}. For the normalization $y^{(0)}=0.0798\,\unit{MeV}^{-\frac{1}{2}}$
 (corresponding to $|C_b|_0=10.0\,\unit{fm}^{-\frac{1}{2}}$)
 was used. 
For each choice of the cut-off $r_c$ the 
effective coupling $h^{(2)}_r$ in this work was chosen such as 
to obtain an optimal description of the $S$-factor. 
The value quoted from Ref.~\cite{Ando:2025ibj} was likewise fitted to the $S$-factor; in this reference a slightly smaller value for the normalization $y^{(0)}=0.058\,\unit{MeV}^{-\frac{1}{2}}$
(corresponding to $|C_b|_0 = 7.27\,\unit{fm}^{-\frac{1}{2}}$) was used. 
}
\label{tab:parE2}
\sisetup{
  round-mode=places,
  scientific-notation = fixed,
}
\begin{tabular}{
  S[round-precision=2, 
  table-format=2.2]
  S[round-precision=2, 
  fixed-exponent = 12, 
  minimum-decimal-digits = 2, 
  table-format=4.2e2]
  S[round-precision=2, 
  fixed-exponent = 12, 
  minimum-decimal-digits = 2, 
  table-format=4.3e2]
  S[round-precision=2, 
  fixed-exponent = 12, 
  minimum-decimal-digits = 2, 
  table-format=4.2e2]
}
\toprule
{$r_c\,[\unit{fm}]$} & \multicolumn{3}{c}{$h^{(2)}_r\,[\unit{MeV}^4]$} \\  
\cline{2-4} & \multicolumn{2}{c}{this work} & \multicolumn{1}{c}{Ref.~\cite{Ando:2025ibj}} \\
\cline{2-3} \cline{3-3}
& "low" & "high" &   \\
\midrule
0.01 & 4.560e12 & 4.591e12 & 4.553e12\\
0.05 & 4.475e12 & 4.5050e12 &  \\ 
0.10 & 4.485e12 & 4.5155e12\\
\bottomrule
\end{tabular}
\end{table}

\begin{figure}[!htb]
  \centering
  \includegraphics[width=0.4\textwidth, trim=50 220 50 70, clip]{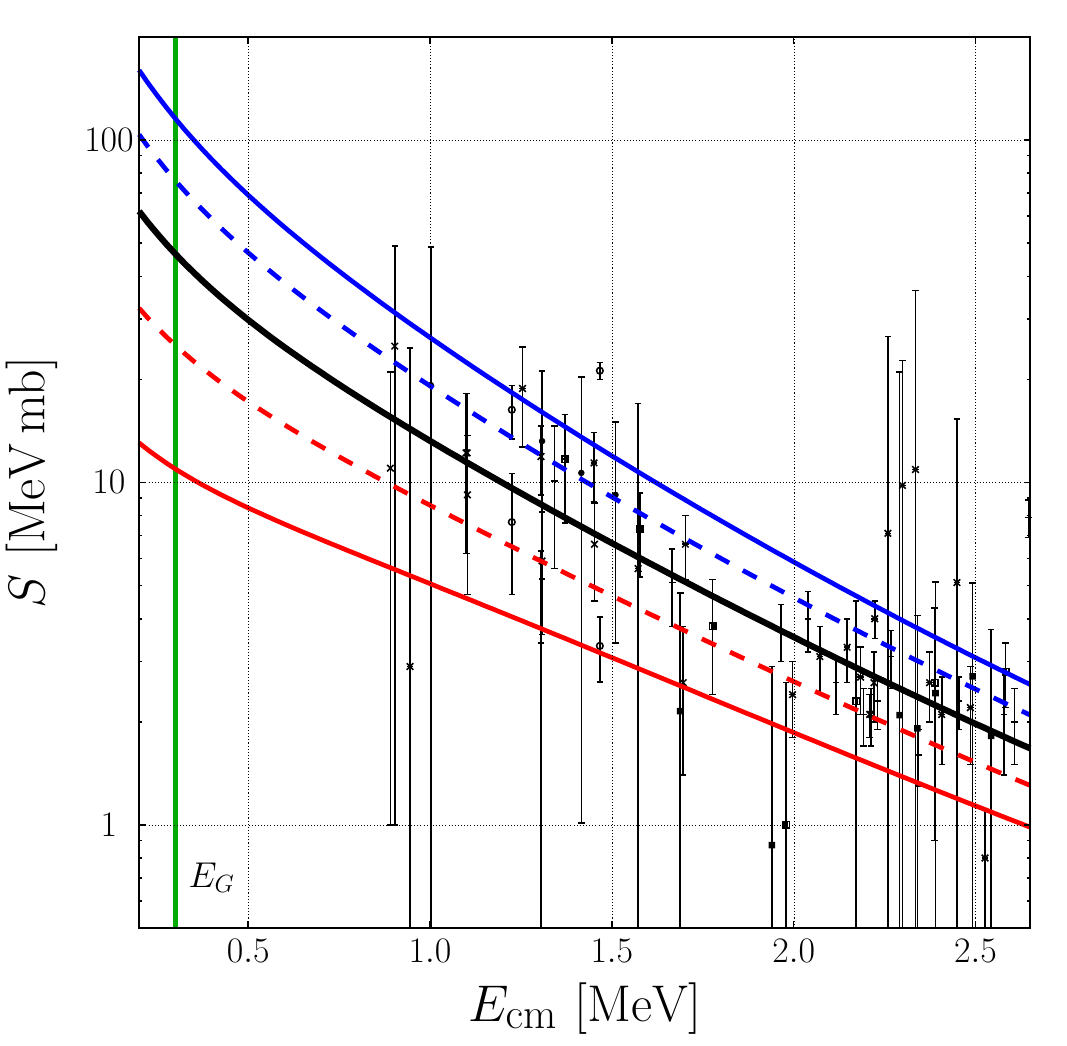}\hspace*{3em}
  \includegraphics[width=0.4\textwidth, trim=50 220 50 70, clip]{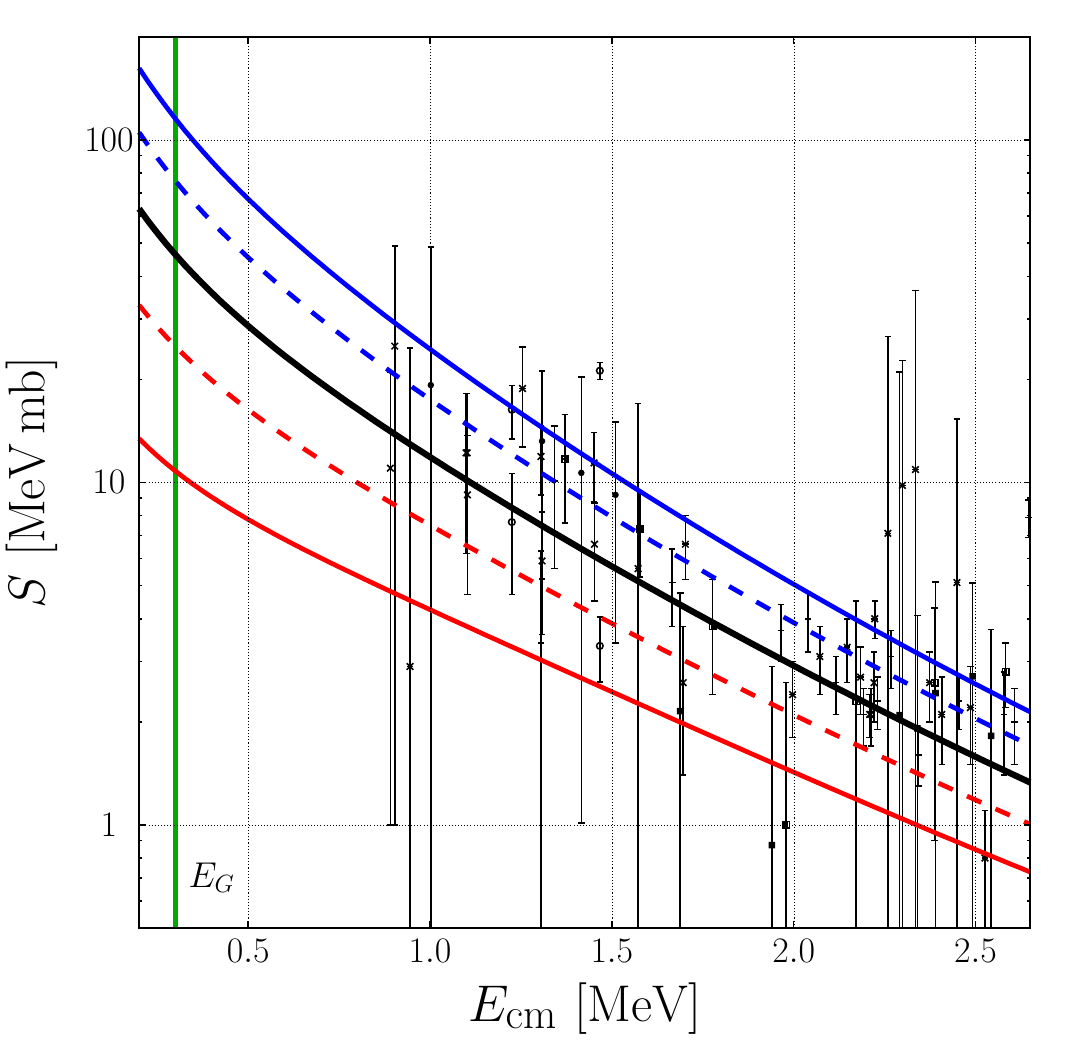}
  \caption{
    Fine structure constant ($\alpha = (1+\delta)\,\alpha_0)$
    variation of the astrophysical $S$-factor of the
    $\nuc{4}{He}+\nuc{12}{C} \to \nuc{16}{O}(2^+) \to \nuc{16}{O}(0^+)$
    radiative capture for two alternative values of the coordinate space cutoff:
    $r_c=0.05\,\unit{fm}$ (left), $r_c=0.10\,\unit{fm}$ (right).
For the parameters, see Table~\ref{tab:parE2}. Also see the caption to Fig.~\ref{fig:SalphavarE2001}.}
  \label{fig:SalphavarE2005010} 
\end{figure} 

Indeed the cut-off dependence is found to be very mild only and thus does not affect the main conclusions presented here, as is also illustrated by the plots of the ratio of the $S$-factor, i.e. $R = S(\alpha)/S(\alpha_0)$ presented in Fig.~\ref{fig:Sratiorc} for the three values of the cut-off $r_c$ discussed here. 

\begin{figure}[!htb]
\includegraphics[width=0.32\textwidth, trim=50 220 50 70, clip]{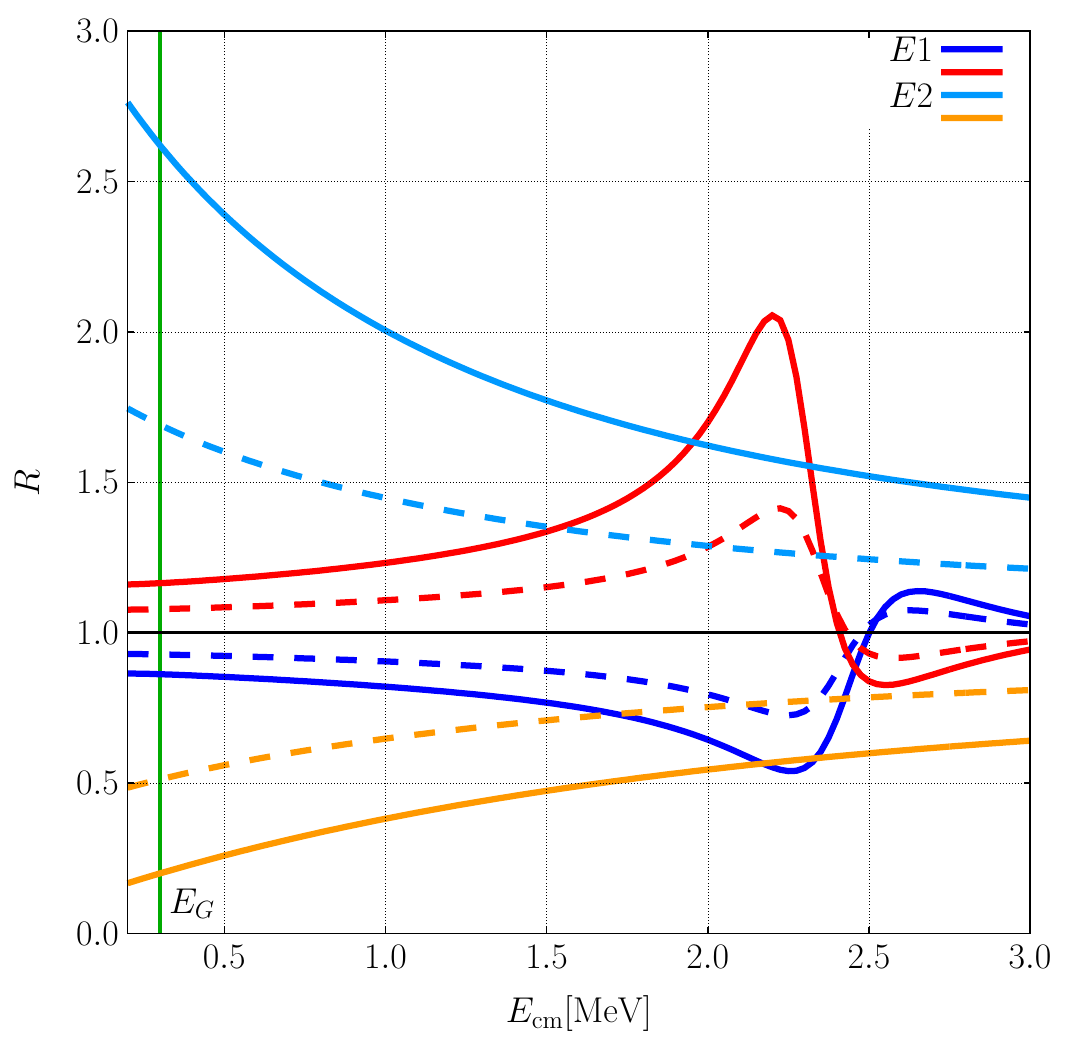}\hfill
\includegraphics[width=0.32\textwidth, trim=50 220 50 70, clip]{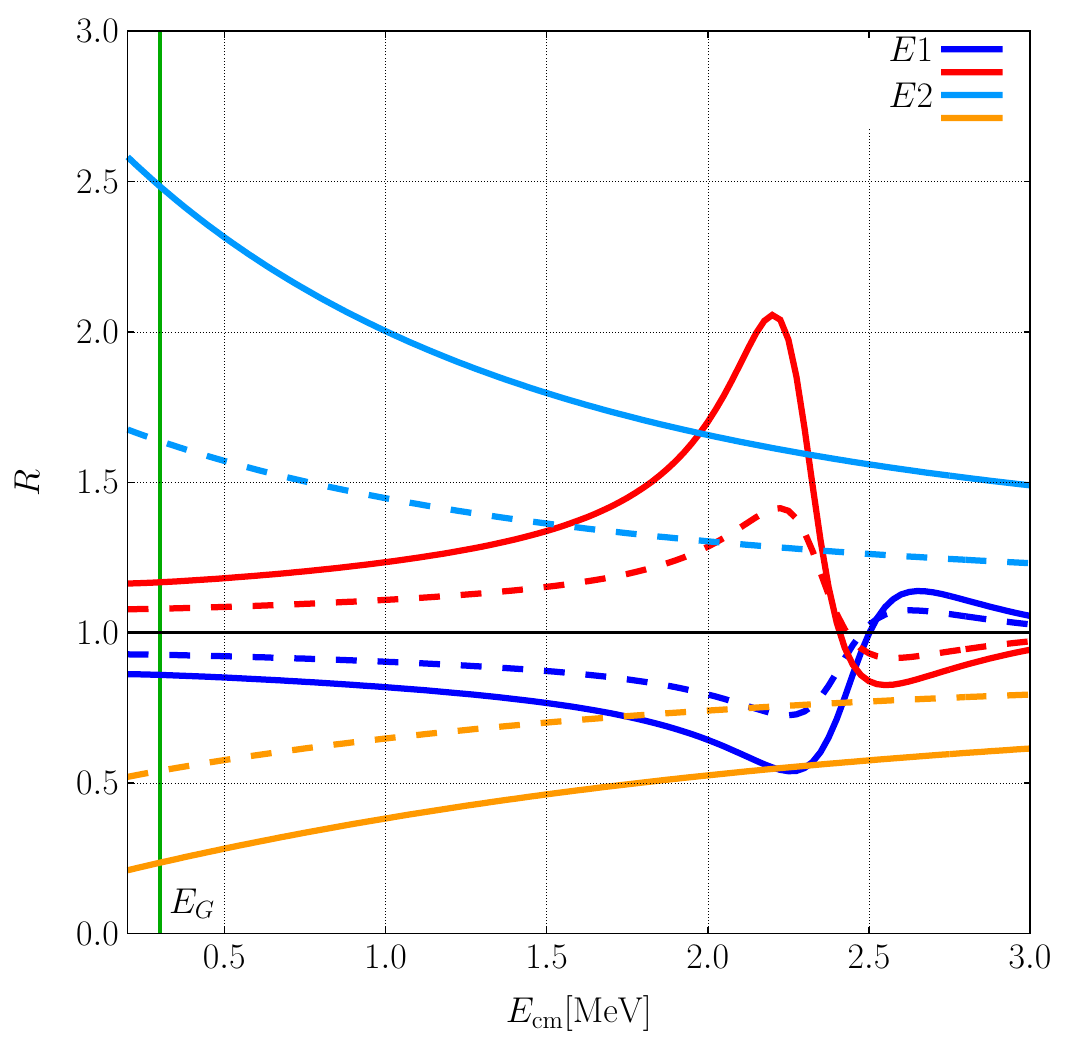}\hfill
\includegraphics[width=0.32\textwidth, trim=50 220 50 70, clip]{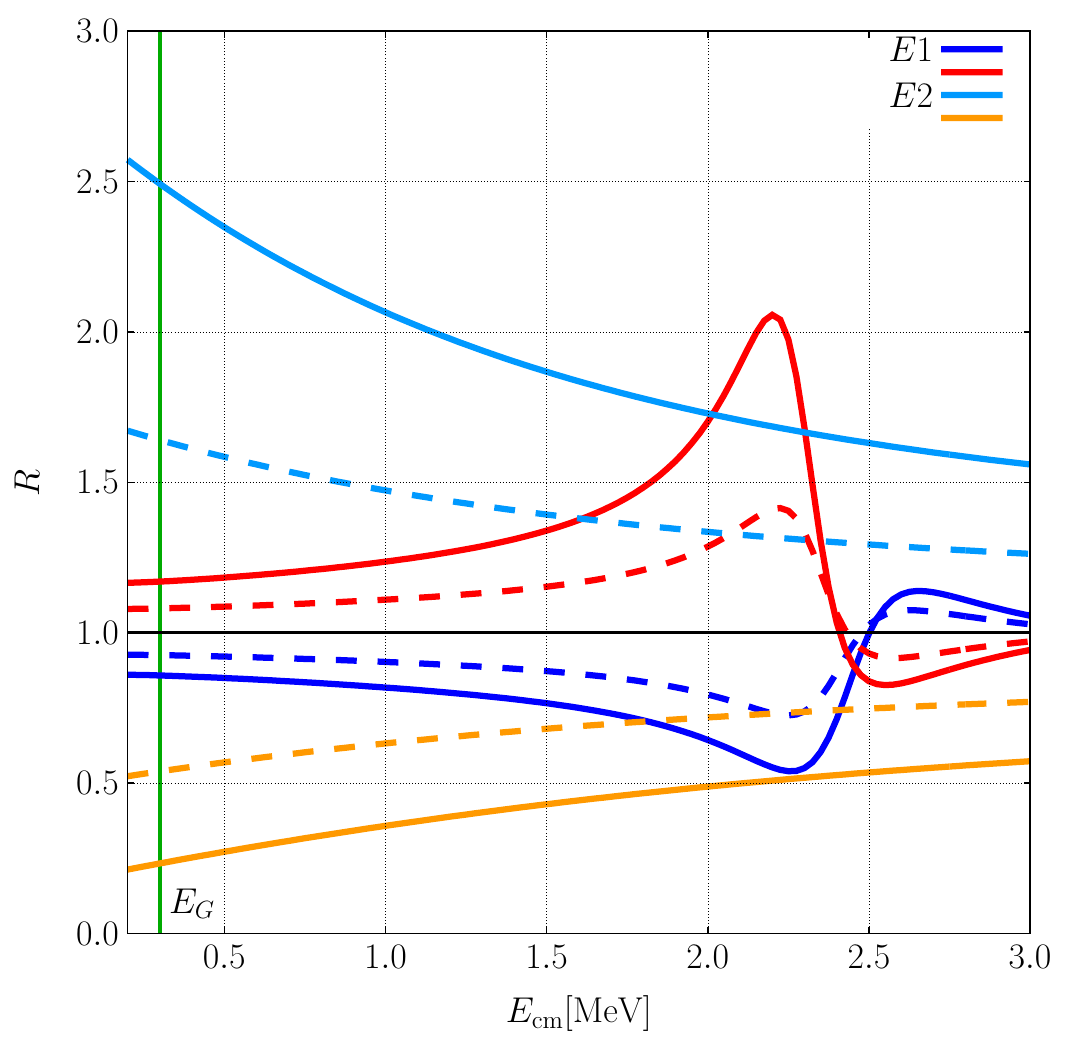}
\caption{
Ratio of the $S$-factor, i.e. $S(\alpha)/S(\alpha_0)$, for $r_c=0.01\,\unit{fm}$ (left), $r_c=0.05\,\unit{fm}$ (middle) and  $r_c=0.10\,\unit{fm}$ (right). Here, $\alpha = \alpha_0\,(1+\delta)$.
For $E1$-radiative capture the red solid (dotted) curve corresponds to $\delta=0.0002 (0.0001)$, the blue solid (dotted) curve to  $\delta=-0.0002 (-0.0001)$ while for $E2$-radiative capture the orange solid (dotted) curve corresponds to $\delta=0.0002 (0.0001)$ and the light blue solid (dotted) curve to  $\delta=-0.0002 (-0.0001)$.
\label{fig:Sratiorc}
}
\end{figure}

\end{widetext}

\bibliographystyle{unsrturl}

\bibliography{references.bib}

\end{document}